\documentclass[letterpaper, 10 pt, conference]{ieeeconf}  %
\IEEEoverridecommandlockouts                              
\IEEEoverridecommandlockouts                         % 
\usepackage{amsmath,amssymb,amsfonts,mathrsfs,amsthm}
\newtheorem{theorem}{Theorem}%[section] % reset theorem numbering for each chapter
\newtheorem{lemma}{Lemma}

\newtheorem{definition}{Definition}
\newtheorem{assumption}{Assumption}
\usepackage{algorithmic}
\usepackage{algorithm}
\usepackage{array}
\usepackage[caption=false,font=normalsize,labelfont=sf,textfont=sf]{subfig}
\usepackage{textcomp}
\usepackage{stfloats}
\usepackage{url}
\usepackage{verbatim}
\usepackage{graphicx}
\usepackage{cite}
\usepackage{color}
\definecolor{winered}{rgb}{0.5,0,0}
\usepackage{scalerel}
\usepackage{xcolor}
\usepackage{mathtools}

\usepackage[colorlinks]{hyperref}
\AtBeginDocument{%
  \hypersetup{
    citecolor=blue,
    linkcolor=red}}

\newcommand{\dgitau}{\mathbf{g}(\x_{t_{i, k}-\tau}, o_{i, t_{i, k}})}
\newcommand{\odit}{o_{i, t_{i, k}}}
\newcommand{\odjt}{o_{j, t_{j, k}}}
\newcommand{\dgit}{\mathbf{g}(\x_{t_{i, k}}, o_{i, t_{i, k}})}

\newcommand{\gi}{\mathbf{g}(\x_{k}, o_{i, k})}

\newcommand{\ek}{\mathbf{e}_k}
\newcommand{\tmax}{\tau_{max}}

\newcommand\x{{\boldsymbol{\theta}}}
\newcommand\xk{{\boldsymbol{\theta}_k}}

\newcommand\xs{{\boldsymbol{\theta}^*}}
\newcommand\xkp{{\boldsymbol{\theta}_{k+1}}}

\newcommand\dx{{\|\xs - \xk\|^2}}
\newcommand\dxpd{{\delta_{k+1}^2}}
\newcommand\dxd{{\delta_k^2}}

\newcommand\vk{\mathbf{v}_k}

\newcommand\okn{o_{i, k}}

\newcommand\bfg{\mathbf{g}}

\newcommand\barg{\bar{\mathbf{g}}}

\newcommand\muiktau{\eta_{k, \tau}^{(i)}}
\newcommand\mudiktau{\eta_{t_{i,k}, \tau}^{(i)}}
\newcommand\mudjktau{\eta_{t_{j,k}, \tau}^{(j)}}

\newcommand{\E}[1]{\mathbb{E}\left[#1\right]}
\newcommand{\eqal}[2]{\begin{equation}\begin{aligned}\label{#1}
			#2
		\end{aligned}\end{equation}}

\newcommand{\inner}[2]{\langle {#1}, {#2} \rangle}
\newcommand{\pars}[1]{\left( {#1} \right)}
\usepackage{cancel}

\begin{document}
\title{\LARGE \bf 
	Finite-Time Analysis of Asynchronous Multi-Agent TD Learning}

\author{Nicol\`o Dal Fabbro, Arman Adibi, Aritra Mitra and George J. Pappas% <-this % stops a space
	\thanks{N. Dal Fabbro is with the Department of Information Engineering, University of Padova. Email: dalfabbron@dei.unipd.it. A. Adibi is with the Department of Electrical and Computer Engineering, Princeton University. Email:  aadibi@princeton.edu. A. Mitra is with the Electrical and Computer Engineering Department, North Carolina State University. Email: amitra2@ncsu.edu. G. J. Pappas is with the Electrical and Systems Engineering Department, University of Pennsylvania. Email: pappasg@seas.upenn.edu. This work was supported by NSF Award 1837253, ARL grant DCIST CRA W911NF-17-2-0181 and the Italian Ministry of Education, University and Research through the PRIN project no. 2017NS9FEY.}% <-
}
\maketitle
\begin{abstract}
Recent research endeavours have theoretically shown the beneficial effect of cooperation in multi-agent reinforcement learning (MARL). In a setting involving $N$ agents, this beneficial effect usually comes in the form of an $N$-fold linear convergence speedup, i.e., a reduction - proportional to $N$ - in the number of iterations required to reach a certain convergence precision. In this paper, we show for the first time that this speedup property also holds for a MARL framework subject to asynchronous delays in the local agents' updates. In particular, we consider a policy evaluation problem in which multiple agents cooperate to evaluate a common policy by communicating with a central aggregator. In this setting, we study the finite-time convergence of \texttt{AsyncMATD}, an asynchronous multi-agent temporal difference (TD) learning algorithm in which agents' local TD update directions are subject to asynchronous bounded delays. Our main contribution is providing a finite-time analysis of \texttt{AsyncMATD}, for which we establish a linear convergence speedup while highlighting the effect of time-varying asynchronous delays on the resulting convergence rate.
% proximity from the desired solution of the FRL problem
%Federated learning (FL) has recently gained much attention due to its effectiveness in speeding up supervised learning tasks under communication and privacy constraints. However, whether similar speedups can be established for reinforcement learning remains much less understood theoretically. Towards this direction, we study a federated policy evaluation problem where agents communicate via a central aggregator to expedite the evaluation of a common policy. To capture typical communication constraints in FL, we consider finite capacity up-link channels that can drop packets based on a Bernoulli erasure model. Given this setting, we propose and analyze \texttt{QFedTD} - a quantized federated temporal difference learning algorithm with linear function approximation. Our main technical contribution is to provide a finite-sample analysis of \texttt{QFedTD} that (i) highlights the effect of quantization and erasures on the convergence rate; and (ii) establishes a linear speedup w.r.t. the number of agents under Markovian sampling. Notably, while different quantization mechanisms and packet drop models have been extensively studied in the FL, distributed optimization, and networked control systems literature, our work is the first to provide a non-asymptotic analysis of their effects in multi-agent and federated reinforcement learning. 
\end{abstract}
\section{Introduction}
Training reinforcement learning (RL) algorithms is well-known to be a critically time consuming task {that can require several data samples to achieve a desired level of accuracy. In an effort to reduce the sample-complexity of RL}, recent research findings in multi-agent RL (MARL) have theoretically shown that, despite temporal correlations in each agent's observations, multi-agent cooperation provides collaborative performance gains \cite{khodadadian2022federated, han}. 
%Training reinforcement learning (RL) algorithms is well-known to be a critically time consuming task. Recent research findings have theoretically shown that, despite the correlated nature of agents' observation processes, multi-agent cooperation  provides collaborative convergence gains \cite{khodadadian2022federated, han}.
%
In a MARL framework where $N$ agents communicate via a central entity - a common setup in the emerging federated RL (FRL) paradigm \cite{qi2021federated} - these gains manifest themselves {in the form of an $N$-fold reduction in the sample-complexity relative to when each agent acts alone}. %In mathematical terms, the convergence speedup means that if when an agent acts alone reaches an approximation error of order $O(\frac{1}{T})$ after $T$ iterations, when $N$ agents act in parallel and cooperate by communicating, they achieve an approximation error of order $O(\frac{1}{NT})$ after $T$ iterations. 
{To achieve such gains, the agents must communicate.}  
However, in real-world multi-agent systems, inter-node communication imposes strong constraints, such as limited link capacity and transmission delays \cite{chen2020joint, koloskova2022sharper, dutta2018slow}. While the effects of these constraints have been widely analyzed for distributed optimization and federated learning (FL), analogous studies in the context of MARL are few and far between. Indeed, collaborative gains under communication constraints have only very recently been established for FRL under finite-rate links and wireless noisy channels~\cite{dal2023federated, dal2023over}.

Given the above premise, our main goal in this paper is to contribute to the literature on \textit{finite-time analyses of MARL under realistic communication models}. To that end, we consider an \emph{asynchronous} MARL  framework in which multiple agents cooperate to evaluate a common policy via temporal difference (TD) learning. To~collaboratively evaluate the common policy, agents transmit their local TD update directions to a central server via up-link communication channels subject to asynchronous bounded delays.
Asynchronous settings of this kind have been theoretically and empirically studied for FL and distributed optimization~\cite{nguyen2022federated, koloskova2022sharper, dutta2018slow}. On the other hand, although asynchronous MARL implementations have shown promising empirical performance, like in the case of parallel actor-learner frameworks \cite{mnih2016asynchronous, nair2015massively}, little to nothing is known regarding their non-asymptotic convergence guarantees and multi-agent collaborative gains. Indeed, the only existing study providing finite-sample convergence guarantees for asynchronous MARL~\cite{AMARL_th} establishes collaborative performance gains only under a simplifying i.i.d. sampling assumption on the agents' observations, i.e., considering observations that are not temporally correlated. However, even in the non-delayed single-agent case, the major technical hurdle in the finite-time analysis of RL algorithms (like TD learning) relative to optimization/supervised learning, comes precisely from the fact that the agent's observation sequence is generated by a Markov chain, and, as such, exhibits temporal correlations. For such settings, finite-time convergence bounds have only recently been provided in \cite{bhandari2018, srikant2019finite} via some fairly involved analysis. Thus, for the MARL setting we consider with Markovian sampling and asynchronous delays, establishing collaborative performance benefits turns out to be highly non-trivial. Nonetheless, providing such an analysis is the main contribution of this paper. 

\textbf{Contributions.} We propose and analyze \texttt{AsyncMATD}, an asynchronous multi-agent TD learning algorithm in which transmission of agents' local TD update directions is subject to asynchronous bounded delays. At each iteration of \texttt{AsyncMATD}, the server
updates the value function associated with the policy to be evaluated using linear function approximation. To do so, it uses potentially stale agents' TD update directions, where the staleness is a consequence of the delays. Our main contribution is to provide a finite-time analysis of \texttt{AsyncMATD} that clearly reveals the effect of the delay sequence on the resulting convergence rate; see Theorem~\ref{th:main}. Remarkably, we establish an $N$-fold linear convergence speedup for \texttt{AsyncMATD} that shows the beneficial effect of collaboration \emph{even in the presence of asynchronous delays in the agents' updates, and Markovian sampling}. %. Notably, our contribution presents the first finite-time analysis for asynchronous MARL, and it jointly establishes collaborative gains 
%a convergence speedup, providing collaborative resulting from multi-agent cooperation. 
%for the first time, theoretically shows the  in MARL \emph{even in the presence of asynchronous delays in the agents' updates}.
%Notably, this paper presents the first finite-time analysis for asynchronous MARL, while jointly establishing a convergence speedup resulting from multi-agent cooperation. 

\section{System Model and Problem Formulation}
\label{sec:system_model}
We consider a setting in which $N$ agents independently interact with replicas of the \textit{same} Markov Decision Process (MDP), which we denote by $\mathcal{M} = (\mathcal{S}, \mathcal{A}, \mathcal{P}, \mathcal{R}, \gamma)$. For the MDP, we consider a finite set $\mathcal{S}$ of $n$ states, a finite action space $\mathcal{A}$, a set of action-dependent Markov transition kernels $\mathcal{P}$, a reward function $\mathcal{R}$, and we denote by $\gamma \in (0,1)$ the discount factor. The goal is to collectively evaluate the value function associated with a  
 policy $\mu:\mathcal{S}\rightarrow\mathcal{A}$. To do so, the agents communicate with a central aggregator (server) via up-link transmissions subject to asynchronous delays. We now review the key concepts regarding policy evaluation with value function approximation. Then, we formally describe the \emph{asynchronous} communication model, outline our key objectives, and highlight the major technical challenges. 

\textbf{Policy Evaluation with Linear Function Approximation.} The purpose of the considered \emph{policy evaluation} setting is to evaluate, for each state $s\in\mathcal{S}$, the value function $\boldsymbol{V}_{\mu}(s)$, which is the discounted expected cumulative reward obtained by playing policy $\mu$ starting from initial state $s$. The policy $\mu$ interacts with the MDP $\mathcal{M}$ to generate a Markov Reward Process (MRP) characterized by a reward function $R_{\mu}:\mathcal{S} \rightarrow \mathbb{R}$, and a Markov chain with transition probability matrix $\mathbb{P}_{\mu}$. Formally, we have
\begin{equation}
	\boldsymbol{V}_{\mu}(s) = \mathbb{E}\left[\sum_{k = 0}^\infty \gamma^kR_{\mu}(s_k)|s_0 = s\right],
 \label{eqn:Value_func}
\end{equation}
where 
$s_k$ represents the state of the Markov chain associated with the kernel $\mathbb{P}_{\mu}$, when initialized from $s_0=s$. The policy evaluation setting  that we consider is \emph{model-free}, i.e., we assume that the Markov transition kernel $\mathbb{P}_{\mu}$ and the reward function $R_{\mu}$ are \textit{unknown}. 

%As such, $V_\mu$ needs to be estimated based on online data, namely, observations of state transitions and rewards acquired by the agents when each of them plays policy $\mu$. 

As is common, we consider a linear value function approximation setting in which the $n$-dimensional value function $\boldsymbol{V}_{\mu}$ is approximated by vectors in a linear subspace of $\mathbb{R}^n$ spanned by a set of $m$ basis vectors $\{\boldsymbol{\phi}_\ell\}_{\ell =1}^m$. Resorting to value function approximation is particularly needed when the state space $\mathcal{S}$ is very large, and indeed usually $m \ll n$.
%In several large-scale practical settings, the size $n$ of the state space $\mathcal{S}$ is large, thereby creating a major computational   challenge. To work around this issue, we will resort to the popular idea of linear function approximation where $\boldsymbol{V}_{\mu}$ is approximated by vectors in a linear subspace of $\mathbb{R}^n$ spanned by a set of $m$ basis vectors $\{\boldsymbol{\phi}_\ell\}_{\ell =1}^m$; importantly,  $m \ll n$. 
Let $\boldsymbol{\Phi} \triangleq [\boldsymbol{\phi}_1, ..., \boldsymbol{\phi}_m] \in \mathbb{R}^{n \times m}$ be the feature matrix. As is standard~\cite{bhandari2018}, we assume that the columns of $\boldsymbol{\Phi}$ are independent and that the rows are normalized. The parametric approximation $\hat{\boldsymbol{V}}_{\boldsymbol{\theta}}$ of $\boldsymbol{V}_\mu$ is given by $\boldsymbol{V}(\boldsymbol{\theta}):=\hat{\boldsymbol{V}}_{\boldsymbol{\theta}} = \boldsymbol{\Phi}\boldsymbol{\theta}$, where $\boldsymbol{\theta} \in \mathbb{R}^m$ is the approximation parameter. Denoting the $s$-th row of $\boldsymbol{\Phi}$ by ${\boldsymbol{{\phi}}}_s'$, the approximation of $\boldsymbol{V}_{\mu}(s)$ is given by $\hat{\boldsymbol{V}}_{\boldsymbol{\theta}}(s) = \langle \x, {\boldsymbol{{\phi}}}_s'\rangle$.

\textbf{Asynchronous multi-agent TD Learning.} In the setting described above, the collective aim of the agents is to estimate the best linear approximation parameter for $\boldsymbol{V}_{\mu}$ in the span of $\boldsymbol{\Phi}$; we denote this optimal parameter by $\boldsymbol{\theta}^*$. To achieve this goal, agents execute a multi-agent TD (MATD) variant of the classical \texttt{TD}(0) algorithm \cite{sutton1988learning}, as we describe next. We assume that agents initialize the MATD algorithm starting from the same state $s_0 \in \mathcal{S}$ and parameter $\boldsymbol{\theta}_0 \in \mathbb{R}^{m}$. In a \textit{synchronous} MATD setting, at each time-step $k \in \mathbb{N}$, a global parameter $\xk$ is broadcasted in down-link by the server to all agents. Each agent $i \in [N]$, in turn, based on the received parameter $\xk$ and its $k$-th observation tuple $o_{i,k} = (s_{i,k}, r_{i,k}, s_{i,k+1})$, computes a local TD update direction $\mathbf{g}(\xk, o_{i,k})$. Note that the observation tuple  $o_{i,k}$ is obtained by playing policy $\mu$ at iteration $k$, which implies taking an action $a_{i,k}=\mu(s_{i,k})$, observing the next state $s_{i,k+1} \sim \mathbb{P}_{\mu}(\cdot| s_{i,k})$, and collecting an instantaneous reward $r_{i,k}=R_{\mu}(s_{i,k})$. Agent $i$'s TD update direction is as follows: 
\begin{equation*}
	\gi = (r_{i,k} + \gamma\langle \boldsymbol{\phi}'_{s_{i,k+1}}, \xk \rangle - \langle \boldsymbol{\phi}'_{s_{i,k}}, \xk \rangle)\boldsymbol{\phi}_{s_{i,k}}'. 
	\label{eqn:localTD}
\end{equation*}

Although all agents play the same policy $\mu$, and interact with replicas of the same MDP, they all experience different realizations of the local observation sequences $\{o_{i,k}\}$. We assume that the local observation sequences $\{o_{i,k}\}$ are \emph{statistically independent} across the agents. 
For each agent $i$, however, the observations over time are correlated since they are all part of a single Markov chain. 

\begin{figure}[t]
\center
\includegraphics[width=1\columnwidth, trim ={5cm 5.5cm 5cm 5cm}, clip]{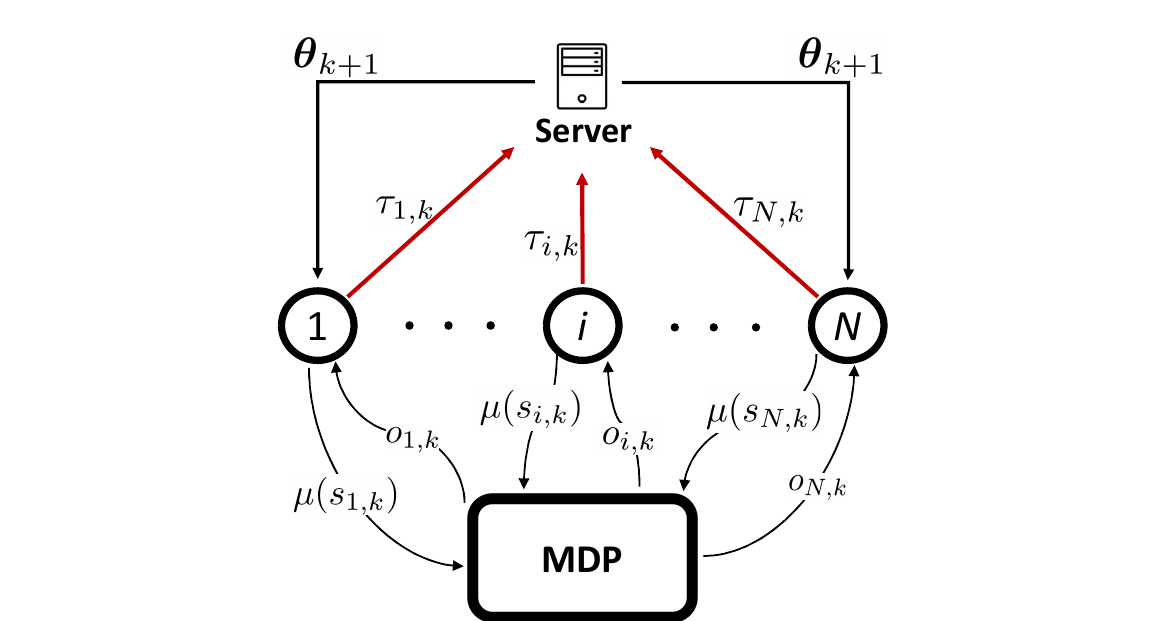}
\caption{System Model. Agents $1, \dots, N$ cooperatively learn a common policy interacting with replicas of the same MDP. At each iteration $k$, the server uses the available delayed update directions with delays $\tau_{1,k}, \dots, \tau_{N,k}$.}
\label{fig:SimConst}
\end{figure}
\textit{Convergence speedup.} Based on the independence property of the observation sequences, one would expect that exchanging agents' local TD update directions should help in reducing the variance in the estimate of $\boldsymbol{\theta}^*$. Achieving this variance reduction in the form of an $N$-fold linear convergence speedup is precisely the aim of multi-agent TD (MATD) learning. The model we described above is a synchronous MARL setting where at each time-step $k$, the server updates the model vector $\xk$ using the average of the agents' local TD update directions from time-step $k$, transmitted over the up-link channels, i.e., \textit{there are no delays}. Departing from this setting, we consider, for the first time, the crucial case in which agents' local TD update directions are computed in an asynchronous fashion. In what follows, we describe our model of asynchrony.

\textit{Asynchrony model of \texttt{AsyncMATD}.} We now describe the model for asynchronous MATD learning that we consider in this paper, which is analogous to the models studied, for example, in FL~\cite{dutta2018slow, nguyen2022federated} and asynchronous MARL\cite{AMARL_th}. At each time-step $k$, the server updates the model vector $\xk$ using the average of asynchronously delayed agents' local TD update directions. Specifically, for each agent $i$, at iteration $k$, the corresponding available TD update direction is subject to a \textit{bounded} delay $\tau_{i, k}$. Define $t_{i,k}\triangleq \left(k-\tau_{i,k}\right)_+$, where, for $x\in\mathbb{R}$, $(x)_+ = \max\{0, x\}$. The server updates the model vector $\xk$ according to the following rule: 
\begin{equation}\label{eq:updateRule}
	\x_{k+1} = \xk + \alpha\mathbf{v}_k,
\end{equation}
where $\alpha$ is a constant step-size/learning rate, and 
\begin{equation}
	\vk = \frac{1}{N}\sum_{i = 1}^{N}\mathbf{g}(\x_{t_{i,k}}, o_{i, t_{i,k}}).
\end{equation}
In this work, we assume that the down-link communication from the server to the agents is not subject to delays. Such an assumption is practically motivated by the fact that in most client-server architectures (e.g., wireless  networks~\cite{chen2020joint}), the main communication bottleneck comes from up-link transmissions, instead of down-link broadcasting. In the rest of the paper, we refer to the update rule in~\eqref{eq:updateRule} as \texttt{AsyncMATD}. Note that the update direction $\vk$ used by the server features iterates $\x_{t_{i,k}}$ and observations $o_{i,t_{i,k}}$ from potentially stale time-steps. Furthermore, the delays $\tau_{1,k}, \dots, \tau_{N,k}$ can differ across agents. 
%In practice, the server could be in the position to always pick the less delayed update direction for each agent, implying that $t_{i,k}$ would be a monotonically increasing sequence with $k$. However, our analysis does not require this to hold, and we only make the following assumption on the delay sequence, which is common in the study of asynchronous optimization and FL~\cite{nguyen2022federated, koloskova2022sharper}.
%
We make the following assumption on the delay sequence, which is common in the study of asynchronous distributed optimization and FL~\cite{nguyen2022federated, koloskova2022sharper}.
\begin{assumption}
	There exists a positive integer $\tau_{max}>0$ such that $0\leq\tau_{i,k}\leq \tau_{max}$, for all $i$ and for all $k$.
\end{assumption}
\textbf{Objective and Challenges.} In the rest of the paper, we provide a \textit{finite-time analysis} of \texttt{AsyncMATD}. This poses several challenges. In fact, even in the single-agent setting, providing a non-asymptotic analysis of \texttt{TD}(0) without performing intermediate projection steps is known to be challenging due to the temporal correlation between the Markov samples in the iterative learning process. Crucially, this challenge is absent in asynchronous stochastic optimization where one assumes i.i.d. data, precluding the use of techniques used in this line of work. For the analysis of \texttt{AsyncMATD}, we encounter further obstacles: the update rule involves the use of multiple \textit{correlated iterates} $\x_{t_{i,k}}$, $i = 1, ..., N$, at which the local TD update directions are asynchronously computed. Indeed, note that, although the observation sequences $o_{i, k}$ are statistically independent across agents, the iterates used to compute the local TD update directions are all correlated. This aspect introduces the need for a much finer analysis when we want to provide finite-time convergence guarantees. Furthermore, unlike a single-agent setting, we aim to establish an $N$-fold linear convergence speedup while jointly dealing with the challenges outlined above. This necessitates a very careful study, which we illustrate below.

\section{Main result}
\label{sec:main_result}
In this section, we state and discuss our main result, which provides a non-asymptotic convergence bound for {\texttt{AsyncMATD}. We start by providing the necessary assumptions and technical machinery. We make the standard assumption that the rewards are uniformly bounded, i.e., that $R_{\mu}(s) \leq \bar{r}, \forall s \in \mathcal{S}$ and for some $\bar{r} > 0$, which ensures that the value function in \eqref{eqn:Value_func} is well-defined. Next, we make a key assumption which is crucial for the non-asymptotic analysis of TD learning algorithms \cite{bhandari2018,srikant2019finite,tsitsiklisroy, khodadadian2022federated, dal2023federated}. 

\begin{assumption}
The Markov chain induced by the policy $\mu$ is aperiodic and irreducible.
\label{assump:mixing}
\end{assumption}

Assumption~\ref{assump:mixing} implies the existence of a unique stationary distribution $\pi$ for the Markov transition matrix $\mathbb{P}_{\mu}$ \cite{levin2017markov}. Let $\mathbf{D}\in\mathbb{R}^{n\times n}$ be a diagonal matrix with entries given by the $n$ elements of $\pi$ and define $\boldsymbol{\Sigma} \triangleq \boldsymbol{\Phi}^\top\mathbf{D}\boldsymbol{\Phi}$. Since $\boldsymbol{\Phi}$ is assumed to be full column rank, $\boldsymbol{\Sigma}$ is full rank with a strictly positive smallest eigenvalue $\omega <1$ that shows up in the convergence analysis. Next, we define the steady-state local TD direction:
{ \begin{equation}
	\barg(\x) \triangleq  \mathbb{E}_{s_{i,k} \sim \pi, s_{i,k+1} \sim \mathbb{P}_{\mu}(\cdot| s_{i,k})}[\mathbf{g}(\x, o_{i,k})], \forall \x \in \mathbb{R}^m. 
\end{equation}}
The \textit{deterministic} recursion $\xkp = \xk + \alpha\barg(\xk)$ captures the limiting behavior of the \texttt{TD}(0) update rule. In \cite{bhandari2018}, it was shown that the iterates generated by this recursion converge exponentially fast to $\boldsymbol{\theta}^*$, where $\xs$ is the unique solution of the projected Bellman equation $\Pi_{\mathbf{D}}\mathcal{T}_{\mu}(\boldsymbol{\Phi}\boldsymbol{\xs})= \boldsymbol{\Phi}\boldsymbol{\xs}$. Here, $\Pi_{\mathbf{D}}(\cdot)$ is the projection operator onto the subspace spanned by $\{\boldsymbol{\phi}_\ell\}_{\ell \in [m]}$ with respect to the inner product $\langle \cdot, \cdot \rangle_{\mathbf{D}},$ and $\mathcal{T}_\mu:\mathbb{R}^{n} \rightarrow \mathbb{R}^{n}$ is the policy-specific Bellman operator \cite{tsitsiklisroy}. Next, we define the notion of mixing time $\tau_{\epsilon}$ that will play a crucial role in our non-asymptotic analysis. 
\begin{definition} \label{def:mix} 
{Let $\tau_{\epsilon}$ be the minimum time step such that} $\Vert \mathbb{E}\left[\mathbf{g}(\boldsymbol{\theta}, o_{i,k})|o_{i,0}\right]-\barg(\boldsymbol{\theta})\Vert \leq \epsilon\left(\Vert \boldsymbol{\theta} \Vert +1 \right), {\forall k \geq \tau_{\epsilon},} \forall \boldsymbol{\theta} \in \mathbb{R}^m, \forall i \in [{N}], {\forall o_{i, 0}}.$\footnote{Unless otherwise specified, we use $\Vert \cdot \Vert$ to denote the Euclidean norm.}
\end{definition}

{ Assumption \ref{assump:mixing} implies that the Markov chain induced by $\mu$ {mixes at a geometric rate \cite{levin2017markov}}, i.e., the total variation distance between $\mathbb{P}\left(s_{i,k}=\cdot|s_{i,0}=s\right)$ and the stationary distribution $\pi$ decays exponentially fast $\forall k \geq 0, \forall i \in [N], \forall s \in \mathcal{S}$. This, in turn, implies the existence of some $K \geq 1$ such that $\tau_{\epsilon}$ in Definition~\ref{def:mix} satisfies $\tau_{\epsilon} \leq K \log(\frac{1}{\epsilon})$~\cite{chenQ}. We infer that the mixing time $\tau_{\epsilon}$ scales logarithmically in the precision dictated by $\epsilon.$
\iffalse
In other words, this means that for a fixed $\boldsymbol{\theta}$, if we want the noisy TD update direction to be $\epsilon$-close (relative to $\boldsymbol{\theta}$) to the steady-state TD direction (where both these directions are evaluated at $\boldsymbol{\theta}$), then the amount of time we need to wait for this to happen scales logarithmically in the precision $\epsilon.$}
\fi
For our purpose, we will set $\epsilon = \alpha^q$, where $q$ is an integer satisfying $q\geq 2$. Unlike the centralized setting where $q=1$ suffices \cite{bhandari2018, srikant2019finite}, to establish the linear speedup property, we will require $q\geq2$. Henceforth, we will drop the subscript of $\epsilon=\alpha^q$ in $\tau_{\epsilon}$ and simply refer to $\tau$ as the mixing time. Let $\dxd \triangleq \dx$ and define by $\sigma \triangleq \max \{1, \bar{r}, \Vert \xs \Vert, \delta_0 \}$ the ``variance" of the observation model for our problem. We can now state our main result. 

\begin{theorem}\label{th:main}
	Consider the update rule of \texttt{AsyncMATD} in (\ref{eq:updateRule}). There exist universal constants {$C_0, C_1, C_2, C_3 \geq 1$}, such that, for $\alpha \leq \frac{\omega(1-\gamma)}{C_0(\tau + \tmax)}$ and $T \geq \tau + 2\tmax$,
\begin{equation}
\label{eqn:main_bound}
\begin{aligned}
%\resizebox{0.98\hsize}{!}{$
	\E{\delta_T^2} 
	&\leq \exp\left(-\frac{\alpha(1-\gamma)\omega T}{2(\tau+\tmax)}\right)C_1\sigma^2 
	\\&+\frac{(\tau + \tmax) \sigma^2}{\omega (1-\gamma)} \left(\frac{C_2\alpha}{N}+C_3 \alpha^3 \right).%$}
\end{aligned}
\end{equation}
\end{theorem}
{ 
\textbf{Discussion:} We now remark on the main takeaways from Theorem \ref{th:main}. From the bound in~\eqref{eqn:main_bound}, we note that \texttt{AsyncMATD} guarantees linear convergence (in mean-square sense) to a ball around $\xs$ whose radius depends on the variance $\sigma^2$ of the noise model. We now comment on the effect of the asynchronous delays on the convergence bound, and on the linear convergence speedup established by the theorem. 

\textit{Effect of asynchronous delays.} From ~\eqref{eqn:main_bound}, note how both the exponent of the linear convergence term and the radius of the noise ball are impacted by the delay sequence via the maximum delay $\tmax$. Indeed, compared to the centralized TD case~\cite[Theorem 7]{srikant2019finite}, and to the synchronous federated TD case~\cite[Theorem 1]{dal2023federated}, we see that for \texttt{AsyncMATD}, the noise ball gets multiplied by the sum of mixing time and maximum delay, i.e.,  $\tau+\tmax$. In essence, our analysis reveals that $\tau+\tmax$ plays the role of an \textit{effective} delay. Interestingly, an immediate implication is that if the underlying Markov chain mixes slowly, i.e., has a larger mixing time $\tau$, then the effect of the delay is less pronounced. This appears to be a novel observation. 

\textit{Linear convergence speedup.}
Compared to the centralized setting \cite[Theorem 7]{srikant2019finite}, the noise variance term in \eqref{eqn:main_bound} gets scaled down by a factor of $N$ up to a higher-order $O(\alpha^3)$ term that, for small enough $\alpha$, is dominated by $(\alpha/N)$. To better illustrate the linear speedup effect, consider the following choice of $\alpha$ and $T$ (define $\bar\tau \triangleq \tau+\tmax$):
\begin{equation}
	\alpha = \frac{\bar{\tau}\log NT}{ \omega (1-\gamma) T}, \hspace{2mm} \textrm{and} \hspace{2mm} T \geq \frac{ 2 C_0 N \bar{\tau}^2 \log NT}{  \omega^2 (1-\gamma)^2}.
	\label{eqn: choices}
\end{equation}
With the above choices, and simple manipulations of the bound in~\eqref{eqn:main_bound}, it can be explicitly shown that 
\begin{equation}
    \mathbb{E}[{\delta_T^2}] \leq O\left(\frac{\sigma^2 \bar\tau^2 \log (NT)}{\omega^2 (1-\gamma)^2 { NT}} \right).
\label{eqn: simplified bound}   
\end{equation}

The above bound tells us that \texttt{AsyncMATD} yields a convergence rate of $O(1/(NT))$, which is a factor of $N$ better than the $O(1/T)$ rate in the centralized case~\cite{bhandari2018}. \emph{We remark that this is the first analysis for asynchronous multi-agent and federated RL that provides finite-time convergence guarantees, while jointly establishing an $N$-fold linear convergence speedup under Markovian sampling.}}
\section{Proof of the main result}
In this section, we prove Theorem \ref{th:main}. We start by introducing the following definitions to lighten the notation:
\begin{equation}\label{eq:dxddifftau}
\begin{aligned}
    \muiktau(\x) & \triangleq \|\E{\mathbf{g}_{i,k}(\x, \okn)|o_{i, k-\tau}} - \barg(\x)\|,\ k\geq \tau, \\
    \delta_{k, h} & \triangleq \|\x_{k} - \x_{k-h}\|,\ k\geq h\geq0,\\
    d_k &\triangleq \max_{k-2\tmax - \tau\leq j\leq k}\E{\delta_{j}^2},\ k\geq \tau + 2\tmax. 
\end{aligned}
\end{equation}
For our analysis, we will need the following result from \cite{bhandari2018}. 
%Specifically, the following Lemma contains the results provided in Lemma 1, 3 and 4 of \cite{bhandari2018}.
\begin{lemma}\label{lemma:bhand}
	The following holds $\forall \boldsymbol{\theta} \in \mathbb{R}^m$: 
\begin{equation*}\label{eq:bhandlemmas}
\begin{aligned}
		\langle \xs - \x, \barg(\x) \rangle \geq \omega(1-\gamma)\|\xs -\x\|^2.
\end{aligned}	
\end{equation*}
\end{lemma}\vspace{0.3cm}
We will also use the fact that the random TD update directions and their steady-state versions are 2-Lipschitz \cite{bhandari2018}, i.e., $\forall i \in [N], \forall k \in \mathbb{N}$, and $\forall \x, \x' \in \mathbb{R}^m$, we have: 
\begin{equation}\label{eq:Lipschitz}
\begin{aligned}
   \max\{\|\bfg(\x) - \bfg(\x')\|, \|\bar{\mathbf{g}}(\x) - \bar{\mathbf{g}}(\x')\| \} \leq 2\|\x - \x'\|.
\end{aligned}
\end{equation}
From \cite{srikant2019finite}, we also have that $\forall i \in [N], \forall k \in \mathbb{N},\forall \x\in \mathbb{R}^m$: 
\begin{equation}\label{eq:boundGradNorm}
    \|\mathbf{g}(\x, \okn)\|\leq 2\|\x\| + 2\bar{r},
\end{equation}
which, squared, using $\bar{r}\leq \sigma$, yields
\begin{equation}\label{eq:bGSq}
	\|\bfg(\x, o_{i,k})\|^2\leq 8(\|\x\|^2 + \sigma^2).
\end{equation}
We will often use the fact that, from the definition of the mixing time in Definition~\ref{def:mix}, we have, for a given iteration $k\geq\tau$, defining $\Theta_{i,k} \triangleq \{\x_{k-\tau}, o_{i,k-\tau}\}$,
\begin{equation}
	\label{eq:mixingProp}
	\Vert\mathbb{E}\left[\mathbf{g}(\boldsymbol{\theta}_{k-\tau}, o_{i,k})|\Theta_{i,k}\right]-\barg(\boldsymbol{\theta}_{k-\tau})\Vert \leq \alpha^{q}\left(\Vert \x_{k-\tau} \Vert +1 \right)
\end{equation}
The proof also relies on the following result from \cite{feyzmahdavian2014delayed}:
\begin{lemma}\label{lemma:OZDAG}
Let ${V_k}$ be non-negative real numbers that satisfy
\begin{align*}
	V_{k+1} &\leq p V_k + q \max_{(k-d(k))_+ \leq \ell \leq k} V_\ell+\beta,
\end{align*}
for $\beta,p,q>0$. Here, $k \geq 0$ and $0 \leq d(k) \leq d_{\max}$ for some $d_{\max}\geq0$. If $p+q<1$, then we have
\begin{align*}
	V_k \leq \rho^k V_0+\epsilon,
\end{align*}
where $\rho = (p+q)^{1/(1+d_{\max})}$ and $\epsilon=\frac{\beta}{1-p-q}$.
\end{lemma}
\noindent We will also use the fact that, for any $a, b \in \mathbb{R}, c\geq0$, 
\begin{equation}\label{eq:halfSq}
ab = a\sqrt{c}\frac{b}{\sqrt{c}} \leq \frac{1}{2}\pars{ca^2 + \frac{b^2}{c}},
\end{equation}
and also the  fact that, for $a_i\in\mathbb{R}, i = 1, ..., N$,
\eqal{eq:sumSqN}{
\pars{\sum_{i = 1}^{N}a_i}^2\leq N\sum_{i = 1}^{N}a_i^2.
}
Equipped with the above basic results, we now provide an outline of our proof before illustrating the technical details.

{\textbf{Outline of the proof}. Recall $t_{i, k} \triangleq \pars{k-\tau_{i,k}}_+$. We write the update rule \eqref{eq:updateRule} as
\eqal{}{
\xkp = \xk  + \alpha\vk = \xk + \alpha\barg(\xk) - \alpha \ek,
}
with $\ek \triangleq \barg(\xk) - \vk$. Thus, 
\eqal{}{
\ek = \frac{1}{N}\sum_{i = 1}^{N}\left(\barg(\xk) - \bfg(\x_{t_{i,k}}, o_{i, t_{i,k}})\right).
}
We analyze the following recursion:
\eqal{}{
\dxpd &= T_1+ \alpha^2T_2-2\alpha T_3 ,\hspace{0.2cm} \textrm{with}\\[4pt]
T_1 &= \|\xk - \xs + \alpha\barg(\xk)\|^2
\\[4pt]
T_2 &= \|\ek\|^2\\[4pt]
T_3 &= \langle \xk - \xs + \alpha \barg(\xk),\ek \rangle
.
}
The most important part of the proof consists of obtaining a bound of the following form:
\eqal{eq:desBound}{
\E{\dxpd} \leq p\E{\dxd} + O(\alpha^2(\tau+\tmax))d_k + B_{\alpha, N},
}
where $d_k$ is as in \eqref{eq:dxddifftau}, $p<1$ is a contraction factor and
\eqal{}{
B_{\alpha, N} = O(\alpha^2(\tau + \tmax))\frac{\sigma^2}{N} + O(\alpha^4)\sigma^2,
}
which guarantees the linear speedup effect with $N$. The bound in \eqref{eq:desBound} allows us to obtain the desired result, by picking a step size small enough and applying Lemma \ref{lemma:OZDAG}. Given this outline, in the following we provide bounds for $\E{T_1}$, $\E{T_2}$ and $\E{T_3}$.

\textit{Bounding $\E{T_1}$.} Note that
\eqal{}{
T_1 &= \|\xk - \xs +\alpha\barg(\xk)\|^2
\\&= \dxd + 2\alpha\langle \xk - \xs,\barg(\xk) \rangle +\alpha^2\|\barg(\xk)\|^2.
}
Note that, using Lemma \ref{lemma:bhand}, we get
\eqal{}{
\langle \xk - \xs,\barg(\xk) \rangle &\leq -(1-\gamma)\omega\dxd,
}
and using~\eqref{eq:Lipschitz} we get
\eqal{}{
\|\barg(\xk)\|^2 &= \|\barg(\xk) - \barg(\xs)\|^2\leq 4\dxd.
}
Combining the two bounds above and taking the expectation,
\eqal{}{
\E{T_1} \leq (1-2\alpha(1-\gamma)\omega)\E{\dxd} + 4\alpha^2\E{\dxd}.
}

\textit{Bounding $\E{T_2}$.} We need the following result.
\begin{lemma}\label{lemma:boundvk}
For $k\geq \tau+ \tmax$, we have
\eqal{}{
\E{\|\vk\|^2}\leq 8\max_{k-\tmax\leq j\leq k}\E{\delta_j^2} + 32\frac{\sigma^2}{N} + 8\sigma^2\alpha^{2q}
}
\begin{proof} We write
\eqal{eq:vkBound}{
\|\vk\|^2
&\leq \frac{2}{N^2}(V_1 + V_2), \hspace{0.3cm} \textrm{with}
\\V_1 & =  \|\sum_{i = 1}^{N}\dgit - \bfg(\xs, o_{i, t_{i,k}})\|^2,
\\V_2 & = \|\sum_{i = 1}^{N}\bfg(\xs, \odit)\|^2.
}
We now bound $V_1$.
\eqal{}{
V_1 &\leq N\sum_{i = 1}^{N}\|\dgit - \bfg(\xs, \odit)\|^2
\\&\overset{\eqref{eq:Lipschitz}}{\leq} 4N\sum_{i = 1}^{N}\delta_{t_{i,k}}^2. \hspace{0.3cm} \textrm{Thus,}
\\ \E{V_1} &\leq 4N\sum_{i = 1}^{N}\E{\delta_{t_{i,k}}^2} \leq 4N^2\max_{k-\tmax\leq j\leq k}\E{\delta_j^2}
}
We now proceed to bound $V_2 = V_{21} + V_{22}$, with
\eqal{}{
V_{21} &= \sum_{i= 1}^{N}\|\bfg(\xs, \odit)\|^2
\\V_{22} &= \sum_{\substack{i,j = 1\\i\neq j}}^{N}\langle \bfg(\xs, \odit),\bfg(\xs, \odjt) \rangle.
}
We see that, using \eqref{eq:bGSq}, we get
\eqal{}{
V_{21}&\leq 8\sum_{i= 1}^{N}(\|\xs\|^2 + \sigma^2)&
\leq 16N\sigma^2.
}
Now, using the fact that the observations $o_{i, k}$ and $o_{j, k'}$ are independent for $i\neq j$ and for any $k, k'\geq 0$,
\eqal{}{
\E{V_{22}} = \sum_{\substack{i,j = 1\\i\neq j}}^{N}&\langle \E{\E{\bfg(\xs, \odit)|o_{i, t_{i,k} - \tau}}},
\\&\E{\E{\bfg(\xs, \odjt)|o_{j, t_{j,k} - \tau}}} \rangle.
} 
Using the fact that $\barg(\xs) = 0$, and Cauchy-Schwarz inequality followed by Jensen's inequality, we can write
\eqal{}{
\E{V_{22}}&\leq \sum_{\substack{i,j = 1\\i\neq j}}^{N}\E{\mudiktau(\xs)}\times\E{\mudjktau(\xs)}
\\&
\leq N^2\alpha^{2q}(\|\xs\| + \sigma)^2\leq 4N^2\alpha^{2q}\sigma^2.
}
Plugging the above bounds on $\E{V_1}$ and $\E{V_2}$ in \eqref{eq:vkBound}, we can conclude the proof of the lemma.
\end{proof}
\end{lemma}
We are now in the position to proceed bounding $\E{T_2}$.
\eqal{}{
\E{T_2} = \E{\|\ek\|^2} &= \E{\|\barg(\xk) - \vk\|^2}
\\& \leq 2\E{\|\barg(\xk)\|^2 + \|\vk\|^2}.
}
Note that $\|\barg(\xk)\|^2 = \|\barg(\xk) - \barg(\xs)\|^2\leq 4\delta_k^2$, and so using Lemma \ref{lemma:boundvk} we get
\eqal{}{
\E{T_2}\leq 24\max_{k-\tmax\leq j\leq k}\E{\delta_j^2} + 64\frac{\sigma^2}{N} + 16\sigma^2\alpha^{2q}.
}

\textit{Bounding $\E{T_3}$.} We now bound $\E{T_3}$, which represents the major technical burden of the proof. 
We need the following result.
\begin{lemma}\label{lemma:delta_kh}
Let $k\geq \tmax + h$, {for some $h\geq0$}. Then,
\eqal{}{
\E{\delta_{k,h}^2}\leq 8\alpha^2h^2\pars{d_k + 4\frac{\sigma^2}{N} + \sigma^2\alpha^{2q}}
}
\begin{proof}Note that, using Lemma~\ref{lemma:boundvk},
\eqal{}{\E{\delta_{k, h}^2} &= \E{\|\xk - \x_{k-h}\|^2}
\\&
\leq h\sum_{l = k-h}^{k-1}\E{\|\x_{l+1} - \x_l\|^2} 
\\&
\leq \alpha^2h\sum_{l = k-h}^{k-1}\E{\|\mathbf{v}_l\|^2}
\\&
\leq \alpha^2h\sum_{l = k-h}^{k-1}(8\max_{l-\tmax\leq j\leq l}\E{\delta_j^2} 
\\&+ 32\frac{\sigma^2}{N} + 8\sigma^2\alpha^{2q})
\\& 
\leq 8\alpha^2h^2\pars{d_k + 4\frac{\sigma^2}{N} + \sigma^2\alpha^{2q}}
.}
\end{proof}
\end{lemma}
Now, we can write $T_3 = K + T_{32}$, with
\eqal{}{
K &= \langle \xk - \xs,\ek \rangle, \\ T_{32} &=  \alpha\langle\barg(\xk), \ek\rangle.
}
Note that, using Cauchy-Schwarz and \eqref{eq:halfSq}, 
\eqal{}{
T_{32}&\leq \frac{\alpha}{2}\left(\|\barg(\xk)\|^2 + \|\ek\|^2\right)
\\&\leq 2\alpha\delta_k^2 + \frac{\alpha}{2}\|\ek\|^2.
}
Taking the expectation and using the bound on $\E{T_2}$,
\eqal{}{
\E{T_{32}}\leq  \alpha(14d_k + 32\frac{\sigma^2}{N} + 8\sigma^2\alpha^{2q}).
}
Now we bound $K$. Define $\barg_{N,k} \triangleq \frac{1}{N}\sum_{i = 1}^{N}\barg(\x_{k-\tau_{i,k}}).$
Adding and subtracting $\barg_{N,k}$, we write 
\eqal{}{
K &= K_1 + K_2, \hspace{0.3cm}\textrm{with}
\\
K_1 &= \inner{\xk - \xs}{\barg(\xk) - \barg_{N,k}},
\\
K_2 &= \inner{\xk-\xs}{\barg_{N,k} - \vk}.
}
Now note that, taking the sum of the second term outside of the inner product in $K_1$, 
\eqal{}{
K_1 &= \frac{1}{N}\sum_{i = 1}^{N}K_{1,i}, \hspace{0.3cm}\textrm{with}
\\K_{1,i} &= \inner{\xk-\xs}{\barg(\xk) - \barg(\x_{t_{i,k}})}.
}
We now bound $\E{K_{1, i}}$. Using \eqref{eq:halfSq} and \eqref{eq:Lipschitz},
\eqal{}{
{K_{1, i}} &\leq {\alpha\tmax}{\delta_k^2} + \frac{1}{4\alpha\tmax}{\|\barg(\xk) - \barg(\x_{t_{i,k}})\|^2}
\\&
\leq {\alpha\tmax}{\delta_k^2} + \frac{1}{\alpha\tmax}\delta_{k, \tau_{i,k}}^2.
}
Note that, using Lemma~\ref{lemma:delta_kh}, which requires $k\geq \tmax + \tau_{i,k}$, which holds for $k\geq 2\tmax$,
\eqal{eq:deltaTau}{
\E{\delta_{k, \tau_{i,k}}^2}\leq 8\alpha^2\tmax^2\pars{d_k + 4\frac{\sigma^2}{N} + \sigma^2\alpha^{2q}}.
}
Taking the expectation and applying \eqref{eq:deltaTau}, we get
\eqal{}{
\E{K_{1,i}}\leq \alpha\tmax \pars{9d_k + 32\frac{\sigma^2}{N} + 8\sigma^2\alpha^{2q}},
}
and note that $\E{K_1}$ is bounded by the same quantity.
We now proceed to bound $\E{K_2}$. 
\eqal{}{
K_2 &= \frac{1}{N}\sum_{i = 1}^{N}K_{2,i}, \hspace{0.3cm}\textrm{with}
\\K_{2,i} &= \inner{\xk-\xs}{\barg(\x_{t_{i,k}}) -\dgit}
\\& = \Delta_{1,i} + \Delta_{2, i} + \Delta_{3, i},\hspace{0.3cm}\textrm{where}
\\[3pt]
\Delta_{1,i} &= \inner{\xk-\xs}{\barg(\x_{t_{i,k}}) - \barg(\x_{t_{i,k} -\tau})},
\\
\Delta_{2,i} &= \inner{\xk-\xs}{\barg(\x_{t_{i,k}-\tau}) - \bfg(\x_{t_{i,k} -\tau}, \odit)}
\\
\Delta_{3,i} &= \inner{\xk-\xs}{\bfg(\x_{t_{i,k} -\tau}, \odit) - \bfg(\x_{t_{i,k}}, \odit)}.
}
Note that, using Cauchy-Schwarz inequality and \eqref{eq:Lipschitz},
\eqal{eq:Delta1i}{
\Delta_{1,i}&\leq \delta_k\|\barg(\x_{t_{i,k}}) - \barg(\x_{t_{i,k}-\tau})\|
\\&
\leq 2\delta_k\delta_{t_{i,k}, \tau}
\\&
\overset{\eqref{eq:halfSq}}\leq \pars{\alpha\tau\delta_k^2 + \frac{\delta_{t_{i,k}, \tau}^2}{\alpha\tau}}.
}
For $k\geq \tau+\tmax$, we can apply Lemma~\ref{lemma:delta_kh} and get
\eqal{}{
\E{\delta_{t_{i,k}, \tau}^2}\leq \alpha^2\tau^2\pars{8d_k + 32\frac{\sigma^2}{N} + 8\sigma^2\alpha^{2q}}.
}
Thus, taking the expectation, we can get
\eqal{}{
\E{\Delta_{1, i}}\leq 9\alpha\tau d_k + 32\alpha\tau\frac{\sigma^2}{N} +8\alpha\tau\sigma^2\alpha^{2q}.
}
Note that, using the Lipschitz property (see~\eqref{eq:Lipschitz}), we get the exact same bound for $\Delta_{3, i}$. Now note that
\eqal{}{
K_2 &= \frac{1}{N}\sum_{i = 1}^{N}\pars{\Delta_{1, i} + \Delta_{3, i}} + \bar{\Delta},\hspace{0.3cm}\textrm{with}
\\
\bar{\Delta} &= \frac{1}{N}\sum_{i = 1}^{N}\Delta_{2, i}.
}
To bound $\bar{\Delta}$, we want to use the geometric mixing property of the Markov chain, that follows from Assumption \ref{assump:mixing}. However, due to the correlations existing between the iterates $\x_{t_{i,k}}$, this needs to be done with special care. Defining $k' \triangleq k-\tmax-\tau$, we start by adding and subtracting $\x_{k'}$ from the first term in the inner product of $\Delta_{2, i}$, getting 
\eqal{}{
\bar\Delta &= \bar{\Delta}_1 + \bar{\Delta}_2, \hspace{0.3cm}\textrm{with}
\\
\bar{\Delta}_1 &= \langle\xk-\x_{k'},\frac{1}{N}\sum_{i = 1}^{N}\barg(\x_{t_{i,k}-\tau}) -\dgitau\rangle,
\\
\bar{\Delta}_2 &= \langle\x_{k'} -\xs,\frac{1}{N}\sum_{i = 1}^{N}\barg(\x_{t_{i,k}-\tau}) -\dgitau\rangle.
}
Note that we can write, using~\eqref{eq:halfSq} and~\eqref{eq:sumSqN},
\eqal{}{
\bar{\Delta}_1&{\leq} \frac{1}{2\alpha\pars{\tmax+\tau}}\delta_{k, \tau+\tmax}^2 \\&+ \frac{\alpha\pars{\tmax+\tau}}{2N^2}\underbrace{\Vert\sum_{i = 1}^{N}\barg(\x_{t_{i,k}-\tau}) -\dgitau\Vert^2}_{G}
}
For $k\geq \tau+2\tmax$, we can apply Lemma~\ref{lemma:delta_kh} and get
\begin{equation}\label{eq:deltaTauMax}
\E{\delta_{k, \tau+\tmax}^2}\leq 8\alpha^2\pars{\tau+\tmax}^2(d_k + 4\frac{\sigma^2}{N} + \sigma^2\alpha^{2q}).
\end{equation}
Now note that
\eqal{}{
G\leq 2\underbrace{\|\sum_{i = 1}^{N}\barg(\x_{t_{i,k}-\tau})\|^2}_{G_{1}} + 2\underbrace{\|\sum_{i = 1}^{N}\dgitau\|^2}_{G_{2}},
}
with $\E{G_{1}}\leq 4N^2\E{\delta_{t_{i,k}-\tau}^2}\leq 4N^2d_k$. Also note that 
\eqal{}{
G_2 &\leq 2G_{21} + 2G_{22},\hspace{0.3cm} \textrm{with}
\\G_{21} & =  \|\sum_{i = 1}^{N}\dgitau - \bfg(\xs, o_{i, k-\tau_{i,k}})\|^2,
\\G_{22} & = \|\sum_{i = 1}^{N}\bfg(\xs, \odit)\|^2.
}
Note that $\E{G_{21}}$ and $\E{G_{22}}$ can be bounded in the same way as $\E{V_1}$ and $\E{V_2}$ in the proof of Lemma~\ref{lemma:boundvk}. We get
\eqal{}{
\E{G_{2}}\leq 8\pars{N^2d_k + 4N\sigma^2 + N^2\sigma^2\alpha^{2q}}.
}
Hence, we get
%\leq 4\alpha\pars{\tmax+\tau}\pars{d_k + 4\frac{\sigma^2}{N} + \sigma^2\alpha^{2q}}\\&+4\alpha\pars{\tmax +\tau}\pars{3d_k + 8\frac{\sigma^2}{N} + \sigma^2\alpha^{2q}}\\&
\eqal{}{
\E{\bar{\Delta}_1}&
\leq 16\alpha\pars{\tmax+\tau}\pars{d_k + 3\frac{\sigma^2}{N} +{\sigma^2\alpha^{2q}}}.
}
Defining $\bar{\Delta}_{2, i}\triangleq\langle\x_{k'} -\xs,\barg(\x_{t_{i,k}-\tau}) -\dgitau\rangle$,
\eqal{}{
\E{\bar{\Delta}_2} = &\frac{1}{N}\sum_{i = 1}^{N}\E{\bar{\Delta}_{2, i}}.
}
Now, defining $\bar{\Theta}_{i, k} \triangleq \{\x_{k'}, \x_{t_{i,k}-\tau}, o_{t_{i,k}-\tau}\}$,
\eqal{}{
\E{\bar{\Delta}_{2, i}} =& \mathbb{E}[\langle\x_{k'} -\xs,\barg(\x_{t_{i,k}-\tau}) -\dgitau\rangle]
\\
= &\mathbb{E}[\langle \x_{k'} -\xs, \\&\barg(\x_{t_{i,k}-\tau})-\E{\dgitau|\bar{\Theta}_{i, k}} \rangle]
\\
\leq& \mathbb{E}[\delta_{k'}
\\&\cdot\underbrace{\|\barg(\x_{t_{i,k}-\tau}) - \E{\dgitau|\bar{\Theta}_{i, k}}\|}_{\bar{\eta}_{k,i}}]
.
}
Now, recall $\Theta_{i, k} = \{\x_{t_{i,k}-\tau}, o_{i, t_{i,k}-\tau}\}$. Note that, for the memoryless property of the Markov chain $\{o_{i,k}\}$, we have
\eqal{}{
\E{\dgitau|\bar{\Theta}_{i, k}} = \E{\dgitau|\Theta_{i, k}}.
}
Indeed, by inspecting the update rule \eqref{eq:updateRule} we see that the parameter $\x_{k'}$ is a function of agent $i$ observations only up to time step $k'-1$, so up to observation $o_{i, k'-1}$. Hence, all the statistical information contained in $\x_{k'}$ has no influence on the random variable $o_{i,t_{i,k}}$ once we condition on $o_{i, t_{i,k}-\tau}$, because $t_{i,k}-\tau = k-\tau_{i,k}-\tau> k'-1= k-\tmax-\tau-1$. Therefore,
\eqal{}{
\bar{\eta}_{k,i} &= \|\barg(\x_{t_{i,k}-\tau}) - \E{\dgitau|{\Theta}_{i, k}}\|
\\&
\overset{\eqref{eq:mixingProp}}\leq \alpha^q\pars{\|\x_{t_{i,k} - \tau}\| + \sigma}.
}
Hence, we get
\eqal{}{
\E{\bar{\Delta}_{2, i}}&\leq \E{\delta_{k'}\alpha^q\pars{\|\x_{t_{i,k} - \tau}\| + \sigma}}
\\&
\leq \E{\delta_{k'}\alpha^q\pars{\delta_{t_{i,k}-\tau} + 2\sigma}}
\\&
\leq \E{\frac{\alpha}{2}\delta_{k'}^2 + {\alpha^{2q-1}}\pars{\delta_{t_{i,k}-\tau}^2 +4\sigma^2}}
\\&
\leq \alpha d_k + 4\alpha\sigma^2\alpha^{2\pars{q-1}}
\\&
\leq \alpha d_k + 4\alpha\sigma^2\alpha^{q}
}
where we used the fact that $\alpha\leq\frac{1}{2}$ and the fact that $2(q-1)\geq q$ for $q\geq2$.
Putting all the above bounds together, and also the fact that $\alpha\leq \frac{1}{16\pars{\tmax+\tau}}$ we get
\eqal{}{
\E{T_3}\leq \alpha\pars{\tmax+\tau}\pars{44d_k + 144\frac{\sigma^2}{N}} + 7\alpha\sigma^2\alpha^q.
}

\textit{Concluding the proof.} Using the bounds on $\E{T_1}, \E{T_2}$ and $\E{T_3}$, we obtain, for $k\geq \tau'\triangleq 2\tmax+\tau$,
\eqal{}{
\E{\dxpd}&\leq (1-2\alpha(1-\gamma)\omega)\E{\dxd}+15\alpha^2\sigma^2\alpha^q\\&+ \alpha^2\pars{\tau+\tmax}\pars{112d_k + 352\frac{\sigma^2}{N}}.
}
We can now write the above bound in the following way:
\eqal{}{
V_{k+1} & \leq pV_k + q\max_{k-\tau'\leq j\leq k}V_j + \beta,\hspace{0.2cm}\textrm{with}
\\V_k &= \E{\dxd},
\\p &= (1-2\alpha(1-\gamma)\omega),
\\q &= 112\alpha^2\pars{\tmax+\tau},
\\\beta &= 352\alpha^2\pars{\tmax+\tau}\pars{\frac{\sigma^2}{N}} + 15\alpha^2\sigma^2\alpha^q.}
Imposing $\alpha\leq \frac{(1-\gamma)\omega}{112(\tau+\tmax)}$, we get $p+q\leq 1 - \alpha(1-\gamma)\omega$, and we can apply Lemma~\ref{lemma:OZDAG} getting
\eqal{eq:aboveBound}{
\E{\delta_T^2}\leq \rho^{T-\tau'}\E{\delta_{\tau'}^2} + \epsilon,
}
with $\rho = \pars{1 - \alpha(1-\gamma)\omega}^{\frac{1}{1+\tau'}}$ and $\epsilon = \frac{\beta}{1-p-q}$. Note that we can easily show that $\rho^{-\tau'}\leq 2$ for $\alpha\leq \frac{1}{16(\tmax + \tau)}$. Furthermore, we can show that $\E{\delta_{\tau'}^2}\leq 3\sigma^2$, as we do next. Note that, using \eqref{eq:bGSq}, 
\eqal{}{
	\|\vk\|^2&\leq \frac{1}{N}\sum_{i = 1}^{N}8(2\delta_{t_{i,k}}^2 + 3\sigma^2)
	\\&
	\leq 16\max_{i = 1, ..., N}\delta_{t_{i,k}}^2 + 24\sigma^2.
}
Using this bound, note that, for any $k\geq 0$
\eqal{eq:formOf}{
	\delta_{k+1}^2 &= \delta_k^2 +2\alpha \langle \x_t - \xs,\vk \rangle + \alpha^2\|\vk\|^2
	\\&
	\overset{\eqref{eq:halfSq}}{\leq }\delta_k^2 + \alpha \delta_k^2 + \alpha\|\vk\|^2 + \alpha^2\|\vk\|^2
	\\&
	\leq (1+\alpha)\delta_k^2 + 2\alpha\|\vk\|^2
	\\&
	\leq (1+\alpha)\delta_k^2 + 32\alpha\max_{i = 1, ..., N}\delta_{t_{i,k}}^2  + 48\alpha \sigma^2,
}
where recall that $t_{i,k}\leq k$. Now define $\bar{p} \triangleq 1+\alpha$, $\bar{q} \triangleq 32\alpha$ and $\nu \triangleq \bar{p}+\bar{q}$ and $\bar{\beta} \triangleq48\alpha \sigma^2$. We now prove by induction that, for all $k\geq 0$, 
\eqal{eq:statAppIndBase}{
	\delta_k^2 \leq \nu^k\delta_{0}^2 + \epsilon_k,
}
where $\epsilon_k = \nu\epsilon_{k-1} + \bar{\beta}$ for $k\geq 1$ and $\epsilon_0 = 0$. The base case is trivially satisfied, because $\delta_0^2 \leq \delta_0^2$. As an inductive step, suppose that~\eqref{eq:statAppIndBase} is true for $0\leq s\leq k$, for some $k\geq 0$, so \eqal{}{
	\delta^2_s\leq \nu^s\delta^2_0 +\epsilon_s, \ \ \ 0\leq s\leq k.
}
Now, we check the property for $k+1$, using~\eqref{eq:formOf}, and noting that $\epsilon_k$ is an increasing sequence:
\eqal{}{
	\delta^2_{k+1} &\leq \bar{p}\delta_k^2 + \bar{q}\max_{i = 1, ..., N}\delta_{t_{i,k}}^2 + \bar{\beta},
	\\&
	\leq \bar{p}(\nu^k\delta^2_0 + \epsilon_k) + \bar{q}\pars{\max_{i = 1, ..., N}\nu^{t_{i,k}}\delta^2_0 + \epsilon_{t_{i,k}}} + \bar{\beta}
	\\&
	\leq \bar{p}(\nu^k\delta^2_0 + \epsilon_k) + \bar{q}(\nu^k\delta^2_0 + \epsilon_k) + \bar{\beta}
	\\&
	\leq (\bar{p}+\bar{q})\nu^k\delta^2_0 + (\bar{p}+\bar{q})\epsilon_k + \bar{\beta}
	\\&
	= \nu^{k+1}\delta^2_0 + \epsilon_{k+1}.
}
From which we can conclude the proof of~\eqref{eq:statAppIndBase}. Now, note that $\epsilon_k = \bar{\beta}\sum_{j = 0}^{k-1}\nu^j$,
and that for $0\leq k\leq {\tau}'$,
\eqal{}{
	\nu^k\leq \nu^{{\tau}'}\leq (1+33\alpha)^{{\tau}'}\leq e^{33\alpha{\tau}'}\leq e^{0.25}\leq 2,
}
imposing $\alpha\leq \frac{1}{132{\tau}'}$. Hence, for $0\leq k\leq {\tau'}$,
\eqal{}{
	{\delta_k^2}&\leq \nu^k\delta_0^2 + \epsilon_k\leq 2\delta_0^2 + \bar{\beta}\sum_{j = 0}^{{\tau'}-1}\nu^j\leq 2\delta_0^2 +2\bar{\beta} {\tau'} \\&= 2\delta_0^2 +2(48\alpha \sigma^2){\tau'}
	\leq 2\delta_0^2 + \sigma^2\leq 3\sigma^2,
}
where we used the fact that $\alpha\leq \frac{1}{100{\tau}'}$ and that $\delta_0^2 \leq \sigma^2.$ We can therefore conclude, writing the bound in \eqref{eq:aboveBound} as
\eqal{}{
\E{\delta_T^2}\leq \exp&\pars{-\frac{\alpha(1-\gamma)\omega T}{2\pars{\tau+\tmax}}}6\sigma^2 \\ &+ 352\frac{\alpha\pars{\tau+\tmax}\sigma^2}{\pars{1-\gamma}\omega N}
+ 15\frac{\alpha^3\sigma^2}{(1-\gamma)\omega}. 
}
\section{Simulations}
In this section, we provide simulation results to validate our theory. We consider an MDP with $|\mathcal{S}| = 100$ states and a feature space spanned by $d = 10$ orthonormal basis vectors; we set the discount factor to $\gamma = 0.5$ and the step size to $\alpha = 0.05$. To simulate the asynchronous delays in the TD update directions, we generate  random delays at each iteration $k$ for each agent $i$, by generating a uniform random variable $\tau_{i,k}$ in the range $[1, \tau_{max}]$. We set $\tau_{max} = 100$. In the multi-agent setting, we set the number of agents to $N = 20$. For each configuration, we plot the average of 20 experiments. It is apparent how the linear speedup property also holds for \texttt{AsyncMATD}. Furthermore, we can see how the gap between the delayed and non-delayed settings decreases in the multi-agent case, compared to the single-agent configuration, by more than one order of magnitude.
\begin{figure}[t]
\center
\includegraphics[width=0.75\columnwidth, trim ={0cm 0cm 0cm 0cm}, clip]{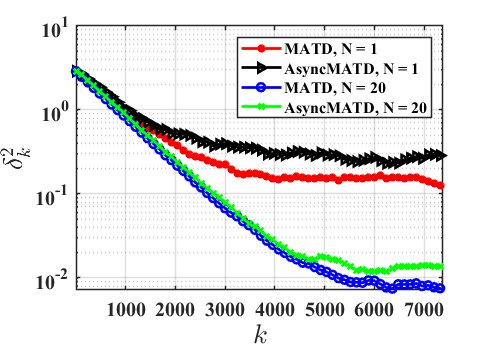}
	\caption{Comparison between vanilla \texttt{MATD} and \texttt{AsyncMATD} in single-agent ($N = 1$) and multi-agent ($N = 20$) settings. For \texttt{AsyncMATD}, we set $\tau_{max} = 100$.}
\label{fig:SimConst}
\end{figure}
\section{Conclusion and Future Work}
We presented and analysed \texttt{AsyncMATD}, providing the first finite-time analysis for asynchronous multi-agent TD learning that jointly establishes a linear convergence speedup with the number of agents. Future work includes more complex delay models, for example considering random and potentially unbounded delays, and the use of delay-adaptive algorithms like the ones proposed for the single-agent and multi-agent settings in~\cite{adibi2024stochastic, fabbro2024dasa}.

\bibliographystyle{IEEEtran}	
\bibliography{IEEEabrv,biblio}

\begin{thebibliography}{10}
\providecommand{\url}[1]{#1}
\csname url@samestyle\endcsname
\providecommand{\newblock}{\relax}
\providecommand{\bibinfo}[2]{#2}
\providecommand{\BIBentrySTDinterwordspacing}{\spaceskip=0pt\relax}
\providecommand{\BIBentryALTinterwordstretchfactor}{4}
\providecommand{\BIBentryALTinterwordspacing}{\spaceskip=\fontdimen2\font plus
\BIBentryALTinterwordstretchfactor\fontdimen3\font minus
  \fontdimen4\font\relax}
\providecommand{\BIBforeignlanguage}[2]{{%
\expandafter\ifx\csname l@#1\endcsname\relax
\typeout{** WARNING: IEEEtran.bst: No hyphenation pattern has been}%
\typeout{** loaded for the language `#1'. Using the pattern for}%
\typeout{** the default language instead.}%
\else
\language=\csname l@#1\endcsname
\fi
#2}}
\providecommand{\BIBdecl}{\relax}
\BIBdecl

\bibitem{khodadadian2022federated}
S.~Khodadadian, P.~Sharma, G.~Joshi, and S.~T. Maguluri, ``Federated
  reinforcement learning: Linear speedup under markovian sampling,'' in
  \emph{ICML}.\hskip 1em plus 0.5em minus 0.4em\relax PMLR, 2022, pp.
  10\,997--11\,057.

\bibitem{han}
H.~Wang, A.~Mitra, H.~Hassani, G.~J. Pappas, and J.~Anderson, ``Federated
  temporal difference learning with linear function approximation under
  environmental heterogeneity,'' \emph{arXiv:2302.02212}, 2023.

\bibitem{qi2021federated}
J.~Qi, Q.~Zhou, L.~Lei, and K.~Zheng, ``Federated reinforcement learning:
  techniques, applications, and open challenges,'' \emph{arXiv preprint
  arXiv:2108.11887}, 2021.

\bibitem{chen2020joint}
M.~Chen, Z.~Yang, W.~Saad, C.~Yin, H.~V. Poor, and S.~Cui, ``A joint learning
  and communications framework for federated learning over wireless networks,''
  \emph{IEEE Transactions on Wireless Communications}, vol.~20, no.~1, pp.
  269--283, 2020.

\bibitem{koloskova2022sharper}
A.~Koloskova, S.~U. Stich, and M.~Jaggi, ``Sharper convergence guarantees for
  asynchronous sgd for distributed and federated learning,'' \emph{Advances in
  Neural Information Processing Systems}, vol.~35, pp. 17\,202--17\,215, 2022.

\bibitem{dutta2018slow}
S.~Dutta, G.~Joshi, S.~Ghosh, P.~Dube, and P.~Nagpurkar, ``Slow and stale
  gradients can win the race: Error-runtime trade-offs in distributed sgd,'' in
  \emph{International conference on artificial intelligence and
  statistics}.\hskip 1em plus 0.5em minus 0.4em\relax PMLR, 2018, pp. 803--812.

\bibitem{dal2023federated}
N.~Dal~Fabbro, A.~Mitra, and G.~J. Pappas, ``Federated {TD} learning over
  finite-rate erasure channels: Linear speedup under markovian sampling,''
  \emph{IEEE Control Systems Letters}, 2023.

\bibitem{dal2023over}
N.~Dal~Fabbro, A.~Mitra, R.~Heath, L.~Schenato, and G.~J. Pappas,
  ``Over-the-air federated {TD} learning,'' in \emph{MLSys23 Workshop on
  Resource-Constrained Learning in Wireless Networks}, 2023.

\bibitem{nguyen2022federated}
J.~Nguyen, K.~Malik, H.~Zhan, A.~Yousefpour, M.~Rabbat, M.~Malek, and D.~Huba,
  ``Federated learning with buffered asynchronous aggregation,'' in
  \emph{International Conference on Artificial Intelligence and
  Statistics}.\hskip 1em plus 0.5em minus 0.4em\relax PMLR, 2022, pp.
  3581--3607.

\bibitem{mnih2016asynchronous}
V.~Mnih, A.~P. Badia, M.~Mirza, A.~Graves, T.~Lillicrap, T.~Harley, D.~Silver,
  and K.~Kavukcuoglu, ``Asynchronous methods for deep reinforcement learning,''
  in \emph{International conference on machine learning}.\hskip 1em plus 0.5em
  minus 0.4em\relax PMLR, 2016, pp. 1928--1937.

\bibitem{nair2015massively}
A.~Nair, P.~Srinivasan, S.~Blackwell, C.~Alcicek, R.~Fearon, A.~De~Maria,
  V.~Panneershelvam, M.~Suleyman, C.~Beattie, S.~Petersen \emph{et~al.},
  ``Massively parallel methods for deep reinforcement learning,'' \emph{arXiv
  preprint arXiv:1507.04296}, 2015.

\bibitem{AMARL_th}
H.~Shen, K.~Zhang, M.~Hong, and T.~Chen, ``Towards understanding asynchronous
  advantage actor-critic: Convergence and linear speedup,'' \emph{IEEE
  Transactions on Signal Processing}, vol.~71, pp. 2579--2594, 2023.

\bibitem{bhandari2018}
J.~Bhandari, D.~Russo, and R.~Singal, ``A finite time analysis of temporal
  difference learning with linear function approximation,'' in \emph{Conference
  on learning theory}.\hskip 1em plus 0.5em minus 0.4em\relax PMLR, 2018, pp.
  1691--1692.

\bibitem{srikant2019finite}
R.~Srikant and L.~Ying, ``Finite-time error bounds for linear stochastic
  approximation and {TD} learning,'' in \emph{Conference on Learning
  Theory}.\hskip 1em plus 0.5em minus 0.4em\relax PMLR, 2019, pp. 2803--2830.

\bibitem{sutton1988learning}
R.~S. Sutton, ``{Learning to predict by the methods of temporal differences},''
  \emph{Machine learning}, vol.~3, no.~1, pp. 9--44, 1988.

\bibitem{tsitsiklisroy}
J.~N. Tsitsiklis and B.~Van~Roy, ``{An analysis of temporal-difference learning
  with function approximation},'' in \emph{IEEE Transactions on Automatic
  Control}, 1997.

\bibitem{levin2017markov}
D.~A. Levin and Y.~Peres, \emph{{Markov chains and mixing times}}.\hskip 1em
  plus 0.5em minus 0.4em\relax American Mathematical Soc., 2017, vol. 107.

\bibitem{chenQ}
Z.~Chen, S.~Zhang, T.~T. Doan, S.~T. Maguluri, and J.-P. Clarke, ``Performance
  of q-learning with linear function approximation: Stability and finite-time
  analysis,'' \emph{arXiv preprint arXiv:1905.11425}, p.~4, 2019.

\bibitem{feyzmahdavian2014delayed}
H.~R. Feyzmahdavian, A.~Aytekin, and M.~Johansson, ``A delayed proximal
  gradient method with linear convergence rate,'' in \emph{2014 IEEE
  international workshop on machine learning for signal processing
  (MLSP)}.\hskip 1em plus 0.5em minus 0.4em\relax IEEE, 2014, pp. 1--6.

\bibitem{adibi2024stochastic}
A.~Adibi, N.~Dal~Fabbro, L.~Schenato, S.~Kulkarni, H.~V. Poor, G.~J. Pappas,
  H.~Hassani, and A.~Mitra, ``Stochastic approximation with delayed updates:
  Finite-time rates under markovian sampling,'' in \emph{International
  Conference on Artificial Intelligence and Statistics}.\hskip 1em plus 0.5em
  minus 0.4em\relax PMLR, 2024, pp. 2746--2754.

\bibitem{fabbro2024dasa}
N.~Dal~Fabbro, A.~Adibi, H.~V. Poor, S.~R. Kulkarni, A.~Mitra, and G.~J.
  Pappas, ``Dasa: Delay-adaptive multi-agent stochastic approximation,''
  \emph{arXiv preprint arXiv:2403.17247}, 2024.

\end{thebibliography}

\end{document}